\documentclass[preprint]{aastex}

\shorttitle{Arp 10: star formation history}
\shortauthors{Bizyaev et al.}

\begin{document}

\title{Modeling the star formation history in the ring galaxy \\
Arp 10 with the help of spectral indices}

\author{Bizyaev D. \altaffilmark{1,2}, Moiseev A. \altaffilmark{3}, 
Vorobyov E. I. \altaffilmark{4,5}}

\altaffiltext{1}{National Optical Astronomy Observatory, Tucson, AZ, USA}

\altaffiltext{2}{Sternberg Astronomical Institute, Moscow, Russia}

\altaffiltext{3}{Special Astrophysical Observatory, Nizhniy Arkhyz, Russia}

\altaffiltext{4}{CITA Fellow} 

\altaffiltext{5}{Dept. of Physics and Astronomy, Univ. of Western Ontario, 
London, Canada}

\begin{abstract}
We use long-slit spectra obtained with the 6-m telescope BTA (Russian
Academy of Sciences) to investigate the history of star formation in the
peculiar ring galaxy Arp 10. The radial distributions of quasi-Lick spectral
indices are calculated from the observed spectra and are compared with the
model spectra, the latter are generated using the Starburst99 population
synthesis code. Our model includes an outward propagating density wave which
triggers star formation in the gas disk, and an old stellar disk. The
metallicity of both old and young stellar populations is a function of
radius. We show that a mix of the young and old populations is required to
explain the radial distribution of the spectral indices. The density wave
propagates outward with a moderate speed (of order 40 km/s), and the
metallicities of both young and old populations decrease with the radius.
The model indices corresponding to the alpha-elements require somewhat
higher metallicities as compared to the Fe-peak elements. We acknowledge
partial support from grant RFBR 04-02-16518.
\end{abstract}


\section{Introduction}

Previous investigations of several ring galaxies have shown that a
considerable part of their stellar population was born at the preset epoch
as a result of the density wave passing. Hence, most of the stellar
population in such objects emerges within a few, or even one event of star
formation. Since the wave triggering the star formation propagates inside
out, we can see stars with different ages at different locations. Unlike the
regular spiral galaxies where stars of different ages are well mixed, the
sites of different ages are separated in space in the ring galaxies. This
fact makes collisional ring galaxies an excellent laboratory for study the
star formation history and general properties of star formation. In this
paper we use two long-slit spectra to constrain models of star formation in
a collisional ring galaxy Arp 10.

\section{Model components and parameters}

\subsection{Propagating Star Formation}

1. An old stellar exponential disk lies in the background. Its surface
brightness and scalelength are parameters of the model which were obtained
from our surface photometry data. The bulge is also taken into account.

2. The young stellar population emerges in the exponential gaseous disk
after triggering of star formation by outwardly propagating ring density
waves. We consider two waves with different speeds which correspond to the
external and internal rings in Arp 10.

3. Metallicity of young population in the external ring is tied to the
observed values (see \citet{Bransford98}), but the radial metallicity gradient is variable
in our models. Metallicity of stellar population in the center of the old
disk and its gradient are the model parameters.

\subsection{Synthetic Spectral Indices}

The metallicities, surface densities, and ages of the stellar populations
are used to find the corresponding spectral indices by interpolation on a
grid of indices.

The grid was computed for a variety of metallicities and ages, and is
obtained from synthetic spectra produced with the Starburst 99 population
synthesis code \citep{sb99}. We got high-res spectra (resolution 0.3A) from
Starburst 99 and convolved them with the instrumental profile of our
observed spectra (resolution 5-8 A). Then the spectral indices were defined
exactly as in \citet{w94} and found for all synthetic spectra (thus we got the
quasi-Lick indices).

\section{Comparison with Observations}

Two long-slit spectra for Arp 10 were taken with the 6-m telescope (Special
Astrophysical Observatory, Russian Academy of Sciences) and the Scorpio
multi-mode spectrograph \citep{scorpio}. Spectral indices, as defined 
by \citet{w94}, and their
uncertainties were obtained from the long-slit spectra at different radii.
SE and NW parts of the long slit spectra were analyzed separately. We
incorporate selected indices to get one common chi-square for them.
The Powell optimization algorithm was applied to minimize this chi-square
value.

\section{Results}

1. The models which take into account only one kind of stellar population
(either young or old) can not explain the radial distributions of spectral
indices.

2. Using spectral indices which correspond to the Fe-peak elements, we got
pretty similar best-fit values for the model parameters with both our
long-slit spectra, see Fig 3 and 4. The mean values of the wave propagation
speed in the region of the outer ring is 35 km/s. The inner ring best-fit
speed, 30 km/s, is very close to the above value. It corresponds to the ages
of the outer and inner rings of 450 and 100 Myr, respectively.

3. The best-fit central metallicity [Fe/H] for the old population is -0.3
dex and its gradient is -0.09 dex/kpc. The gradient of metallicity of young
population is about -0.05 dex/kpc. In comparison to the SE parts of our
long-slit spectra, the NW parts give rather similar best-fit model
parameters with a smaller value for the outer ring speed.

4. The results shown above reveal two major recent events of propagating
star formation in Arp 10 induced by the passing of a satellite-intruder (the
satellite was unambiguously identified in \citet{bmv04}).

This research will be continued with a more realistic physical modeling of
Arp 10 (enhancing \citet{vb03}) and with more constrains from studies of gas
kinematics and abundances analysis.

\acknowledgments

This project was partially supported by grant of the Russian Foundation for
Basic Research, RFBR 04-02-16518.


\begin{figure}
\plotone{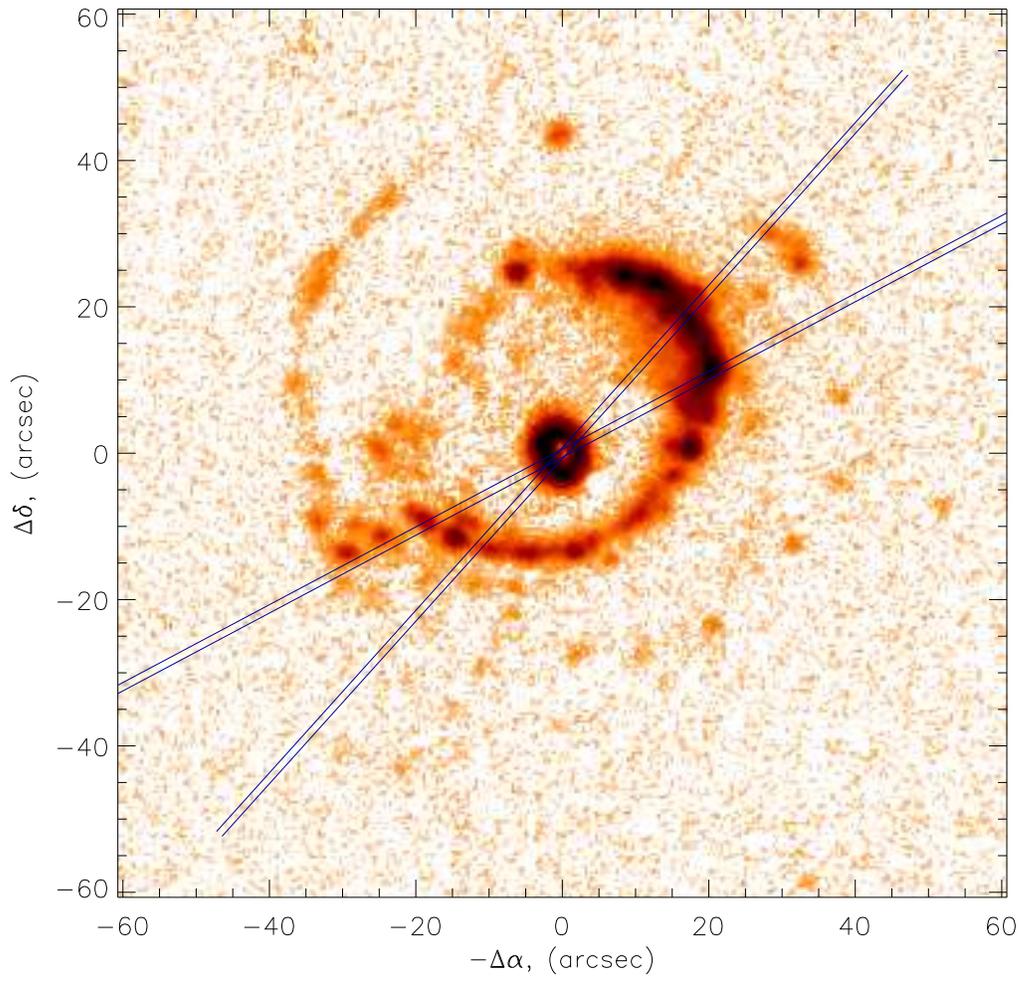}
\caption{Image of Arp 10. Positions of the slits are shown in the figure.
\label{fig1}}
\end{figure}

\begin{figure}
\plotone{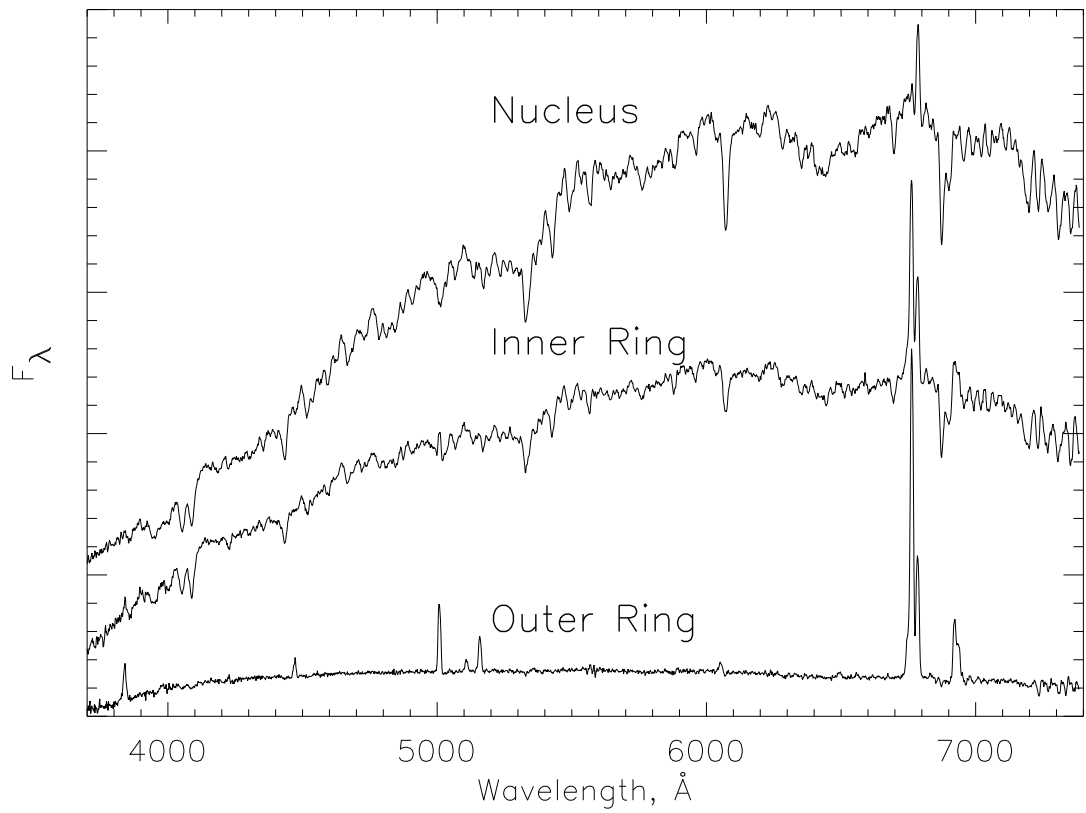}
\caption{Long-slit spectra of the nucleus, inner ring, and outer ring
in Arp 10.
\label{fig2}}
\end{figure}

\begin{figure}
\plotone{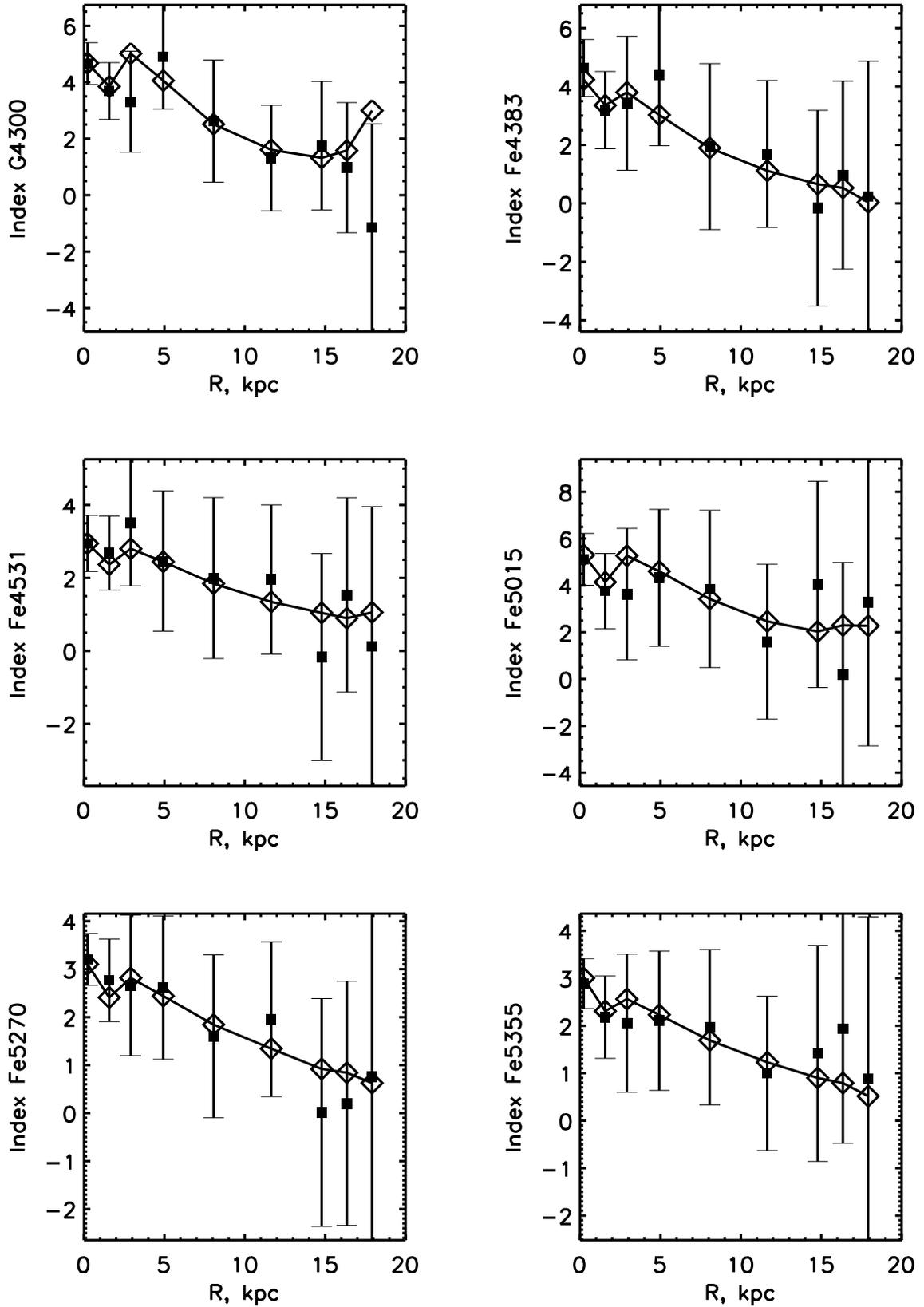}
\caption{Radial distribution of best-fit synthetic spectral indices (diamonds
and solid curve) for spectra \#1. Observed indices are shown by filled
squares with the error bars.
\label{fig3}}
\end{figure}

\begin{figure}
\plotone{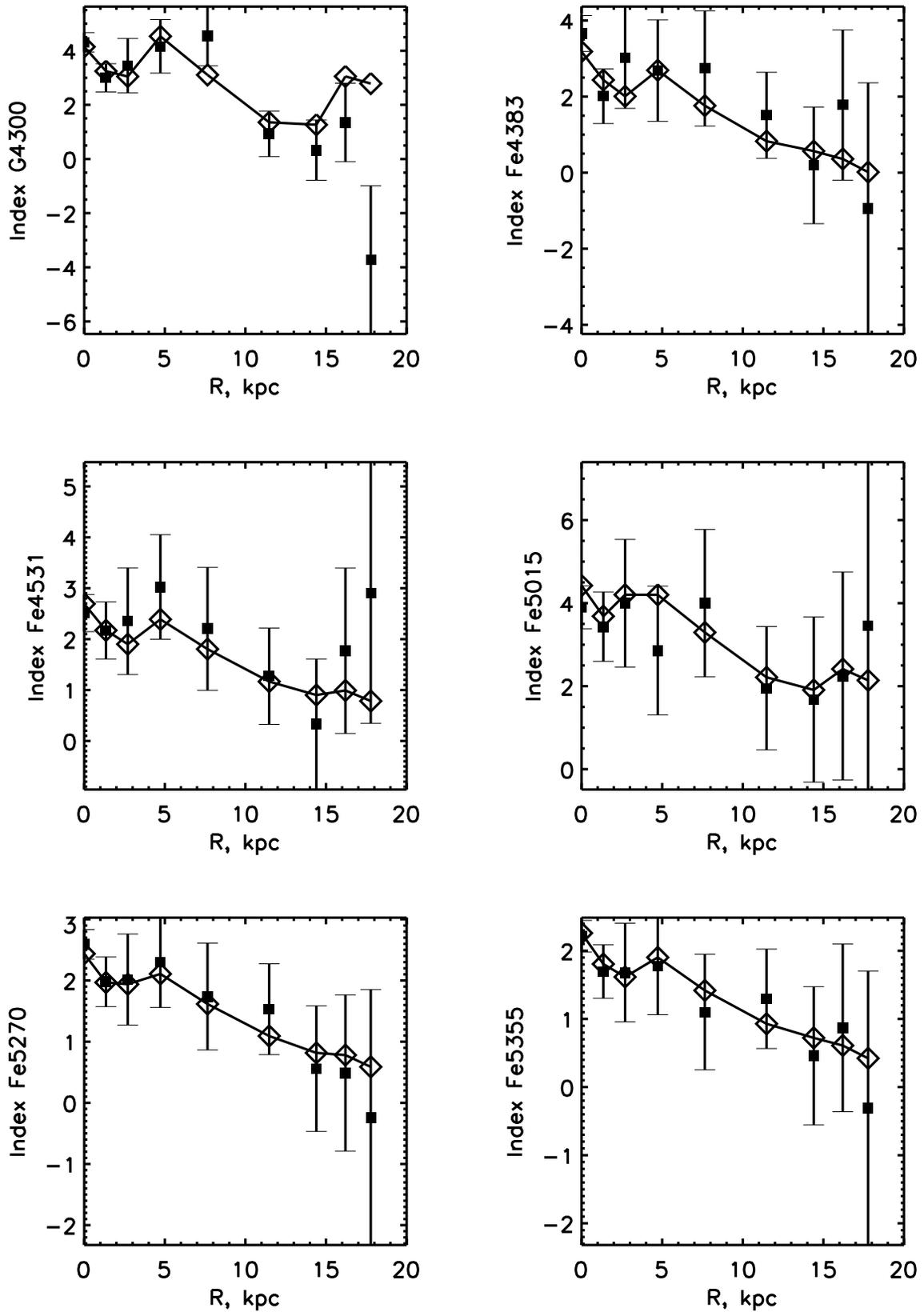}
\caption{The same as in Fig. 3 for the spectra \#2.
\label{fig1}}
\end{figure}

\end{document}